%% file: PersistenceMitigation.tex
\documentclass[12pt]{article}

\usepackage{fancyhdr}
\usepackage{aaspp4}
\usepackage{epsf}
\usepackage{flushrt}
\usepackage{graphicx, afterpage}
\usepackage{listings}
\usepackage{color}
\usepackage{colortbl}
\usepackage{float}
\usepackage{indentfirst}
\usepackage{subfigure}
\usepackage[format=plain, labelsep=none, justification=justified, font=footnotesize, labelfont=it, singlelinecheck=false]{caption}

\usepackage{natbib}
\usepackage[colorlinks=true,citecolor=blue,urlcolor=cyan]{hyperref}
\input{kbsmacs.tex}
\reservedcharson
\actcharon

\usepackage{geometry}
\def\ssection#1{\section{\hbox to \hsize{\large\bf #1\hfill}}}
\def\ssectionstar#1{\section*{\hbox to \hsize{\large\bf #1\hfill}}}
\def\ssubsectionstar#1{\subsection*{\hbox to \hsize{\normalsize\bf #1\hfill}}}

\long\def\symbolfootnote[#1]#2{\begingroup%
  \def\thefootnote{\fnsymbol{footnote}}\footnote[#1]{#2}\endgroup%
  \def\footnoterule{\null}}

\definecolor{gray}{gray}{0.85}

\begin{document}

% Logo
\begin{figure*}[h]
\begin{minipage}[t]{40cm}
\includegraphics[height=35mm]{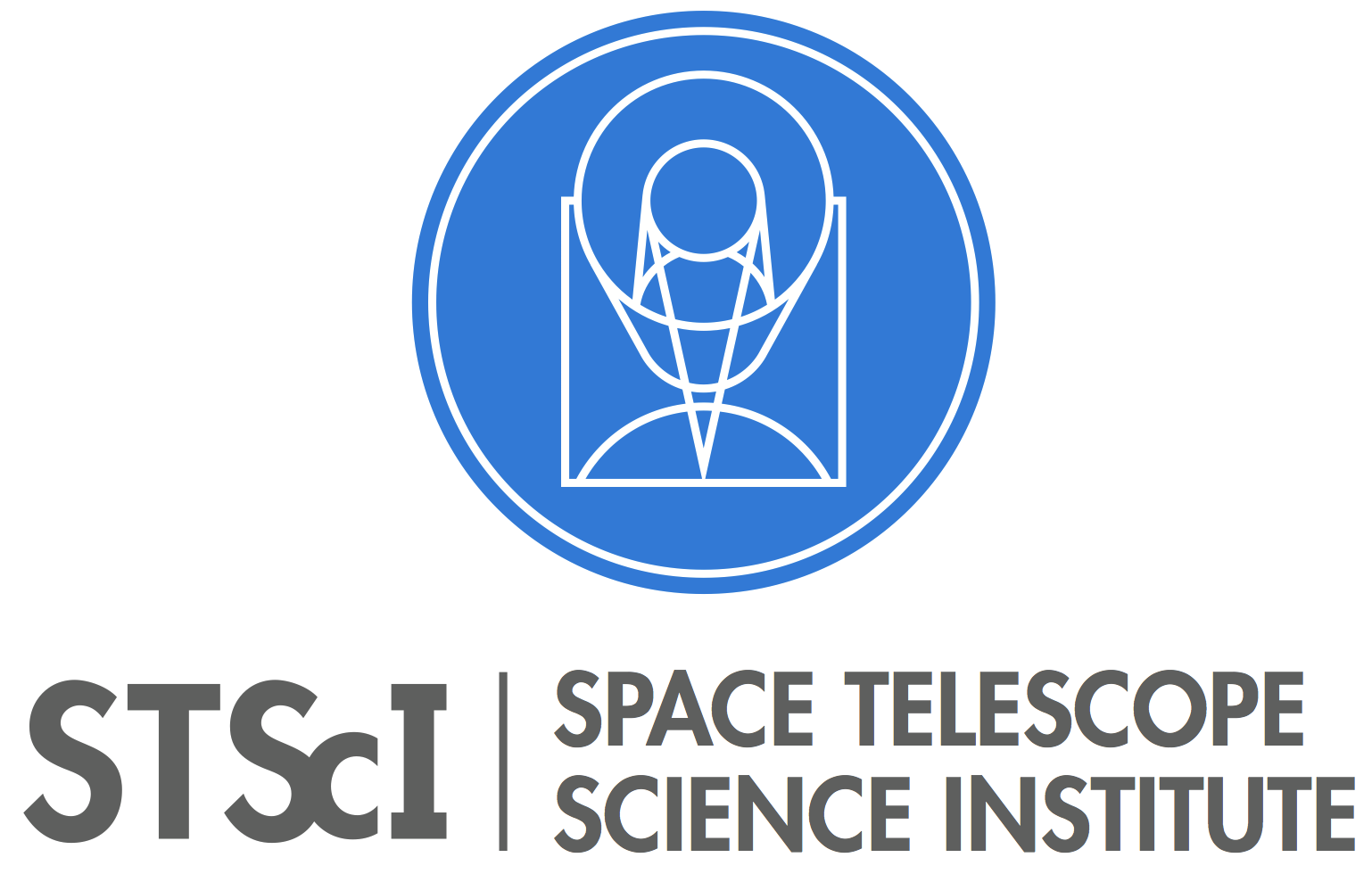}
\end{minipage}
\end{figure*}

% ISR Number
\begin{flushright}
\vskip -1.3truecm
{\bf Instrument Science Report WFC3 2019-13}
\end{flushright}

% Title and authors
\begin{flushright}
%{\huge\bf \hfill Persistence Mitigation of WFC3/IR Spatial Scan Observations}
{\huge\bf \hfill Pre-Flashing WFC3/IR Time-Series, Spatial Scan Observations}
\rule{145mm}{0.3mm}
\smallskip \\
 \hfill {K.~B. Stevenson \& W. Eck}\\
 September 30, 2019
 %\today
 \end{flushright}
 \medskip

%%%%%%%%%%%%%%%%%%%%%%%%%%%%%%%%%%%%%%%%%%%%%%%%%%%%%%%%%%%%%%%%%%%%%%%%%%%%%%%%%%%%%%%%%%%
%                                      Abstract
%%%%%%%%%%%%%%%%%%%%%%%%%%%%%%%%%%%%%%%%%%%%%%%%%%%%%%%%%%%%%%%%%%%%%%%%%%%%%%%%%%%%%%%%%%%

\hrule height 1.5pt
\smallskip
\noindent \large{\bf A}\footnotesize{\bf BSTRACT}

\normalsize\noindent{\textit{
 Spatial scan observations using WFC3's IR channel exhibit time-dependent systematics (in the form of a ramp or hook) that have been attributed to the effects of persistence.  The amplitude of these systematics is often two orders of magnitude larger than the signal sizes of interest and, therefore, must be carefully modelled and removed.   The goal of this calibration program (CAL-15400) is to mitigate these systematics by continuously illuminating the detector while repeatedly reading it out during Earth occultation (termed pre-flashing).  Compared to standard observations, we are able to reduce the amplitude of the systematic effect by a factor of $\sim$7 (from 1.30\% to -0.19\%), thus confirming our hypothesis that the detector more quickly reaches an equilibrium state when subjected to higher flux levels.  Compared to the latest modeling techniques \citep{Zhou2017}, we achieve a marginal improvement in the white light curve precision ($\Delta$rms = -8{\pm}9~ppm); therefore, pre-flashing is an equally effective means to mitigate WFC3's instrument systematics.  We conclude that pre-flashing does not warrant future consideration due to the increase in the number of channel select mechanism (CSM) motions, effort required to implement, and equivalent ability to model instrument systematics with current techniques.
 %We estimate that implementing pre-flashing for all future time-series observations in spatial scan mode would result in a 15\% increase in the number of moves by WFC3's channel select mechanism (CSM) and would recover an average of 30 {\hst} orbits per year.  We conclude that, although pre-flashing may not be advisable for the WFC3 instrument, it may be a viable strategy for similar near-infrared {\jwst} instruments if they exhibit a similar behavior.
 %We also consider an alternative strategy in which pre-flashing is used only once at the beginning of each visit would yield a 3\% increase in the number of CSM moves; however, future calibration work would be needed to validate this strategy's benefits.
 %In theory, the use of pre-flashing could improve {\hst}'s efficiency by including the first {\hst} orbit in the analyses of time-series observations.  However, future implementation of our exact strategy may pose a long-term risk to the health and safety of the WFC3 instrument; therefore, we consider several other strategies that may yield comparable results with less risk.   Future calibration work is needed to test these strategies and develop a viable path forward.
}}

\smallskip
\medskip
\hrule height 1.5pt

\symbolfootnote[0]{Copyright {\copyright} 2019 The Association of
  Universities for Research in Astronomy, Inc. All Rights Reserved.}

%%%%%%%%%%%%%%%%%%%%%%%%%%%%%%%%%%%%%%%%%%%%%%%%%%%%%%%%%%%%%%%%%%%%%%%%%%%%%%%%%%%%%%%%%%%
%                                   Introduction
%%%%%%%%%%%%%%%%%%%%%%%%%%%%%%%%%%%%%%%%%%%%%%%%%%%%%%%%%%%%%%%%%%%%%%%%%%%%%%%%%%%%%%%%%%%

\ssectionstar{Introduction}
\normalsize{

%Reference first spatial scan observations
In 2011, the {\em Hubble Space Telescope} ({\hst}) performed its first Wide Field Camera 3 (WFC3) time-series, spatial scan observations.  This includes a test observation of GJ~1214 from program 12325 (PI: MacKenty) and two transiting exoplanet observations from a general observer program (GO-12181, PI: Deming).  In the eight years since, WFC3 has made over 200 time-series, spatial scan observations, mostly targeting exoplanets as they pass in front of or behind their parent stars and fast-rotating brown dwarfs.

During this time, the standard observing strategy has been to acquire an additional orbit of data at the start of each time-series observation as a means to stabilize the detector.  By excluding the first orbit, the systematics from all subsequent orbits within that observation exhibit a repeatable behavior.  The repeatability of {\hst}'s orbit-dependent systematics (often referred to as the ``ramp'') is key to fitting time-series observations and achieving $<100$~ppm precision.  This ramp is typically attributed to charge trapping in the detector pixels.

%Cite Zhou paper to model effect
Recent work by \citet{Zhou2017} has introduced a new, physically-motivated model based on charge-trapping theories that fits the ramp in all {\hst} orbits.  They find that the scatter in the residuals from the first {\hst} orbit are consistent with those from subsequent orbits (within a given observation) and conclude that data from the first orbit no longer need be discarded.  Out of precaution and to improve the baseline flux constraint, the WFC3 team continues to support the acquisition of a stabilizing orbit at the start of each visit.
%Community response to this technique has been mixed and most (if not all) WFC3 time-series observations continue to acquire a stabilizing orbit at the start of each visit.

%Discuss Knox Long's previous work (2014-14)
\citet{ISR-2014-14} attempt to mitigate charge trapping by pre-conditioning the detector during Earth occultation (CAL-13573, PI: Long).
They found that by filling detector pixels with light from the tungsten lamp and holding the charge for $\sim$42 minutes before reading out the electronics, \citet{ISR-2014-14} were able to reduce the amplitude of the effect by a factor of $\sim$2.
They also note that science observations with a larger number of scans per orbit have a shorter exponential timescale and, thus, stabilize faster.  They conclude that each time a pixel is illuminated, a fraction of the available traps are filled irrespective of how long the pixel maintained its fluence level.

%The purpose of this program... hypothesis
In reviewing their work, we adopt an alternative interpretation, namely that the detector stabilizes faster due to higher flux levels.  We note that brighter targets require faster scan rates and, as a result, contain more scans per orbit; therefore, the two parameters are closely linked.  We use this interpretation for the basis of our calibration program.  The hypothesis is that by pre-flashing the detector with a bright source (i.e. high flux) while repeatedly reading out the electronics, the detector will more quickly reach its equilibrium state.  Performing this strategy within each Earth occultation would then mitigate the ramp seen during science observations.

In the sections below, we provide the details of our observations, data reduction, and analyses; present our findings from this calibration program; and consider the impact of implementing the pre-flashing strategy on future time-series, spatial scan observations.
%potential implementation strategies within the confines of not impacting the long-term health and safety of the instrument.

%\citet{ISR-2012-08} discuss considerations for preparing WFC3 spatial scan observations
%Discuss magnitude of effect

}

%%%%%%%%%%%%%%%%%%%%%%%%%%%%%%%%%%%%%%%%%%%%%%%%%%%%%%%%%%%%%%%%%%%%%%%%%%%%%%%%%%%%%%%%%%%
%                                   Observations
%%%%%%%%%%%%%%%%%%%%%%%%%%%%%%%%%%%%%%%%%%%%%%%%%%%%%%%%%%%%%%%%%%%%%%%%%%%%%%%%%%%%%%%%%%%

\ssectionstar{Observation Strategy}

In this calibration program (CAL-15400; PI: Stevenson), we perform time-series observations of a point source in spatial scan mode in coordination with Tungsten lamp pre-flash observations during Earth occultation.
We select GJ~1214 as our point source because we need a reference system that was previously observed without pre-flashing, thus providing a baseline for comparison.  GJ~1214 is a relatively quiet, slowly-rotating M dwarf and the transiting planet in this system has no detectable thermal emission at these wavelengths, which would have made our analysis more challenging.

We apply the same observing strategy as the final 10 visits from GO program 13021 (PI: Bean).  Namely, we obtain 103 second exposures with the G141 grism (NSAMP = 15; SAMP-SEQ = SPARS10) in the 256$\times$256 subarray mode.  Exposures scan perpendicular to the dispersion direction at a rate of 0.12 {\arcsec}/sec and alternate directions (i.e., round trip mode).  During test building for the calendar we find that Visits 02 -- 04 are too long (as implemented by Program 13021) because they did not allow enough time for gyro bias updates. Unlike the original program, bias updates are now necessary to mitigate pointing failures.  To recover $\sim$5 minutes at the end of each orbit for gyro bias updates, we scheduled three fewer scans per {\hst} orbit (16 vs 19).  Similar to GO-13021, we obtain a direct image of GJ~1214 at the start of each {\hst} orbit using the F130N filter  (NSAMP = 3; SAMP-SEQ = RAPID).  Table \ref{tab:obs} summarizes the observations from both programs.  Visit 14 represents a typical observation from GO-13021 and is representative of time-series, spatial scan observations in general.

Unlike GO-13021, we precede each science orbit with our programmed persistence mitigation strategy. This involved exposing the WFC3/IR detector to light emitted by the on-board Tungsten lamp during Earth occultation.  Each of 55 exposures lasts 22.3 seconds (NSAMP = 4; SAMP-SEQ = SPARS10).  The goal is to achieve a median fluence comparable to that of GO-13021 (21,860 e\sp{-}/pixel).  Using the G141 grism, we reach a median fluence of 19,230 e\sp{-}/pixel.  This corresponds to a median flux of 862 e\sp{-}/s.  See Appendix A for addition information on how this observation was scheduled.

%tab%%%%-------------------------------------------------------%%%%%%
\begin{table}[t]
\makebox[\textwidth][c]{
\begin{tabular}{ccccc}
\hline
Label       & Program ID    & Visit \#  & Obs. Date     & Median Fluence  \\
            &               &           &               & (e\sp{-}/pixel)   \\
\hline
Pre-Flashed & 15400         & 01-04     & 2018-08-10    & 19,230{\pm}50\\
Standard    & 13021         & 14        & 2013-07-06    & 21,860{\pm}1300\\
\hline
\end{tabular}
}
\caption{\label{tab:obs}
\textsl{Observation summary.}}
\end{table}
%tab%%%%-------------------------------------------------------%%%%%%

%%%%%%%%%%%%%%%%%%%%%%%%%%%%%%%%%%%%%%%%%%%%%%%%%%%%%%%%%%%%%%%%%%%%%%%%%%%%%%%%%%%%%%%%%%%
%                                   Data Reduction
%%%%%%%%%%%%%%%%%%%%%%%%%%%%%%%%%%%%%%%%%%%%%%%%%%%%%%%%%%%%%%%%%%%%%%%%%%%%%%%%%%%%%%%%%%%

\ssectionstar{Data Reduction and Analysis}
%\ssectionstar{Data Reduction and Analysis Using QuickLook's Spatial Scan Monitor}

To reduce the GJ~1214 data, we used software that is part of the WFC3 Quicklook project \citep{Bourque2017} and is intended to monitor time-series observations that use the spatial scan mode \citep{ISR-2019-SSMonitor}.
Briefly, the Spatial Scan Monitor is used to track data quality over time and quantify how it varies with certain parameters (e.g., pointing drift, fluence, etc.).  %The code is fully automated, which has the benefit of yielding uniform results with no human interaction.  %However, at times it may yield unexpected results when data have been acquired using non-standard observing techniques.  These troublesome datasets are easily identified and removed when performing statistical studies.
The data reduction software makes use of the IMA files rather than the FLT files because the former often yield more robust results.  The reduced data consist of band-integrated light curves (flux vs.~time) with auxiliary information.
%how we reduce lamp data
No reduction is required for the tungsten lamp data.  We use the FLT frames and compute median flux values over the same subarray region used to analyze the GJ~1214 data.
%LC fitting
%Zhou et al, Stevenson et al
We fit the white light curves using standard techniques \citep[e.g., ][]{Stevenson2014a}.  For both datasets, we adopt a quadratic correction in time and a custom-written version of the ramp model described by \citet{Zhou2017}.

%%%%%%%%%%%%%%%%%%%%%%%%%%%%%%%%%%%%%%%%%%%%%%%%%%%%%%%%%%%%%%%%%%%%%%%%%%%%%%%%%%%%%%%%%%%
%                                   Results
%%%%%%%%%%%%%%%%%%%%%%%%%%%%%%%%%%%%%%%%%%%%%%%%%%%%%%%%%%%%%%%%%%%%%%%%%%%%%%%%%%%%%%%%%%%

\ssectionstar{Results}
\normalsize{

\ssubsectionstar{Pre-Flashed Data}
%Measured fluence during Earth occultation (when Tungsten lamp is on) exhibits unexpected behavior (see figure)
%   For each HST orbit, the measured fluence ramps up quickly and slowly decays
%   I suspect this could be the lamp itself (want to talk to Mario when he returns)
We measure the flux during Earth occultation (when the Tungsten lamp is turned on) to assess its stability.  Figure \ref{fig:tungsten} shows a sharp increase in flux at the beginning of each orbit followed by a steady decline.  Tests indicate that the observed trend is consistent across the entire detector, not just the science region exposed by GJ~1214.  %FINDME
We consider whether the observed trend is the result of
%The observed trend could be the result of 
variations in the brightness of the tungsten lamp, charge trapping in the detector, or both.  The tungsten lamp has a warm-up period of at least 40 seconds before subsequent commanding can start; however, despite this delay, data from the UVIS Bowtie Monitor program \citep[CAL 11808, 11908, 12344, 12688, 13104, and 13072;][]{ISR-2013-09} shows that the first tungsten lamp exposure is consistently lower than the third exposure ($>200$ seconds later) by a few percent.  We note that the two channels (UVIS and IR) use physically different, but identically-produced lamps.  This information suggests that the tungsten lamp (at least partially) contributes to the ramp shown in Figure~\ref{fig:tungsten}.  A future calibration program is necessary to fully quantify the lamp's contribution to this ramp.

%fig%%%%-------------------------------------------------------%%%%%%
\begin{figure}[tp!]
\begin{center}
\makebox[\textwidth][c]{\includegraphics[width=0.8\linewidth]{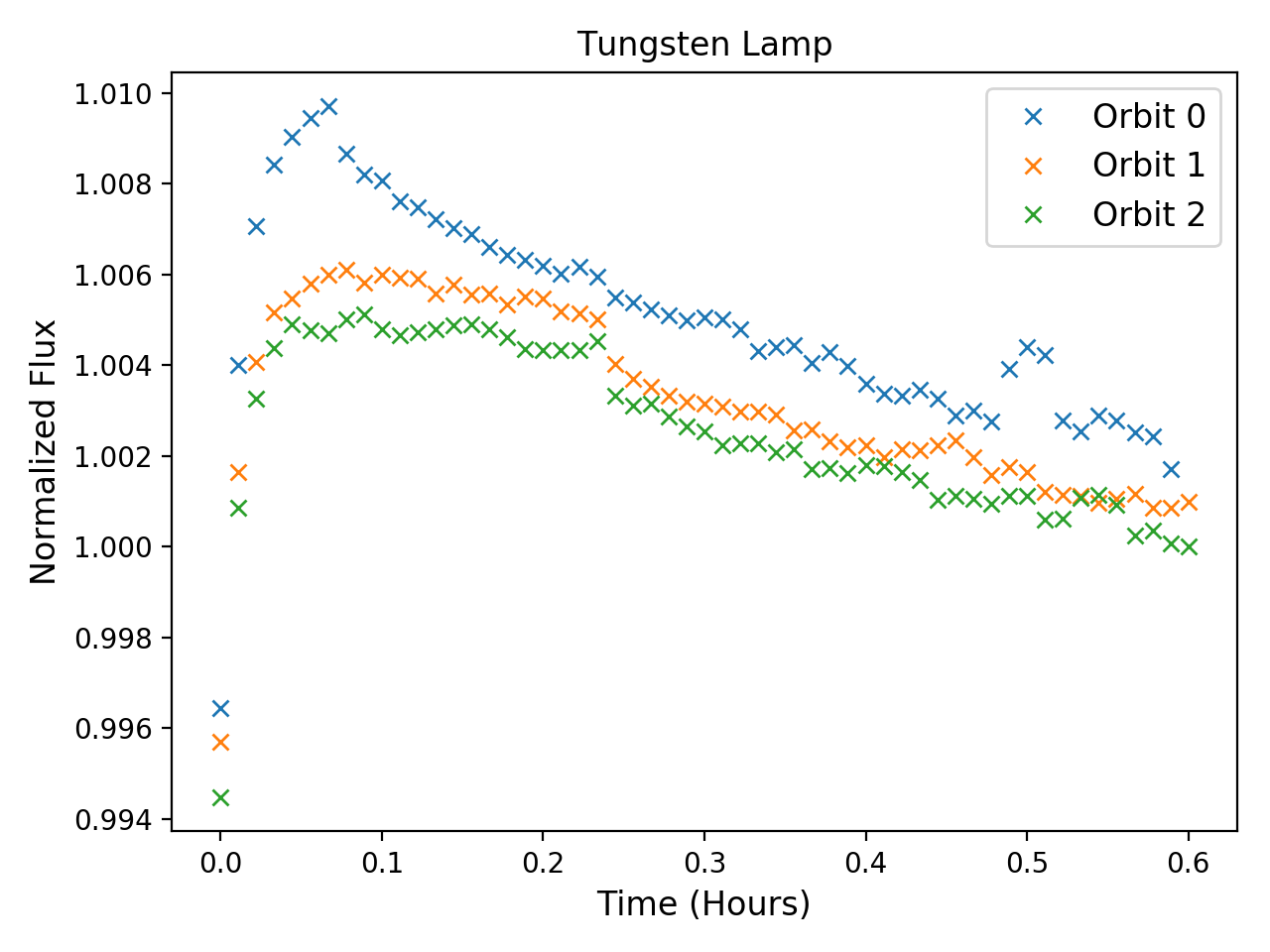}}\vspace{-1em}
\caption{\label{fig:tungsten}
\textsl{Measured flux from the tungsten lamp over three orbits.  The flux is normalized to the last measurement in Orbit 2.  The shape of the measured flux variation is consistent between orbits, but its cause is unclear.  %FINDME We leave a more in-depth study of these data for future work.
}}
\makebox[\textwidth][c]{\includegraphics[width=0.8\linewidth]{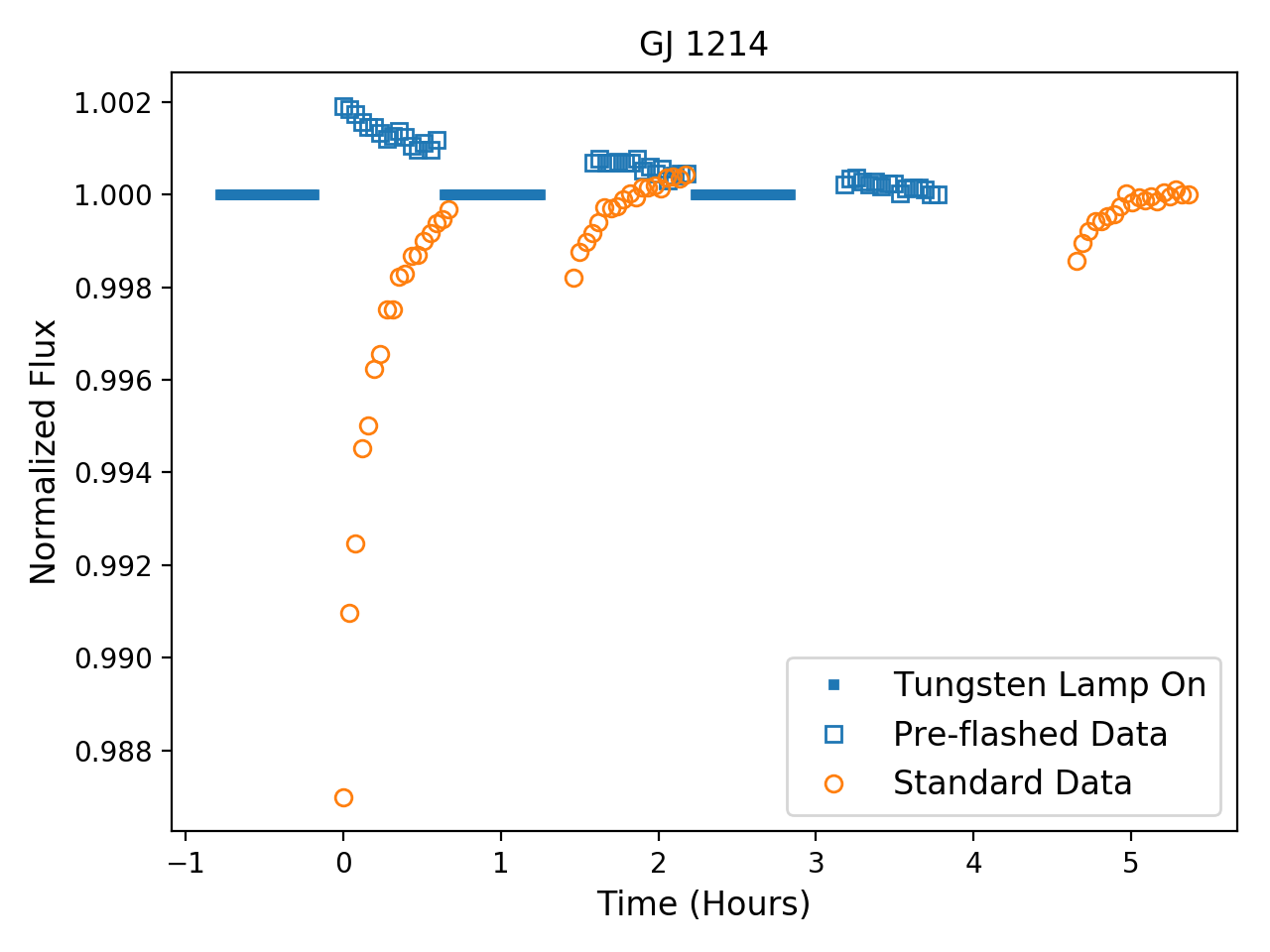}}\vspace{-1em}
\caption{\label{fig:lc}
\textsl{Normalized white light curves with and without pre-flashing the detector.  The pre-flashed data (blue squares) yield a less-pronounced ramp ({\hst} orbit-dependent systematic) relative to the standard dataset (orange circles).  The improvement is the result of illuminating and cycling the detector during Earth occultation (blue lines).  [Note: we have removed the third orbit from the standard dataset because it contains the transit of GJ~1214b.]
}}
\end{center}
\end{figure}
%\afterpage{\clearpage}
%fig%%%%-------------------------------------------------------%%%%%%

\ssubsectionstar{Persistence Mitigation}
Figure \ref{fig:lc} depicts the normalized flux as a function of time (the white light curves) from both programs.  Data acquired using the standard method (GO-13021) exhibit significant ramps, with flux increases of 1.26\%, 0.22\%, and 0.14\% over orbits 1, 2, and 4, respectively.  The total increase from the first to fourth orbit is 1.30\%.  By comparison, data acquired with the pre-flashing technique exhibit smaller decrements in flux (-0.09\%, -0.02\%, and -0.02\% over orbits 1, 2, and 3, respectively).  The total change over all three orbits is -0.19\%.  The factor of 13 reduction in the persistence systematic within the first orbit suggests that, with pre-flashing, it may be possible to retain this orbit for science purposes.
Additionally, after the first orbit the pre-flashed data exhibit minimal curvature in the persistence systematic.  %With more testing, it may be possible to calibrate the tungsten lamp brightness or exposure duration to further reduce the amplitude of this systematic.

\ssubsectionstar{Light Curve Residuals}

After fitting each white light curve with a quadratic function in time and a ramp model similar to that of \citet{Zhou2017}, we compute the residual rms values.  Figure \ref{fig:res} plots a histogram of the residuals.  We find that the pre-flashed data exhibit less scatter than the standard data (64 vs 72 ppm rms); however, the standard error on the residuals is {\pm}9~ppm, so the difference is not statistically significant.  For comparison, \citet{Kreidberg2014} report an rms of 70 ppm for their example observation.  The expected photon-limited rms is $\sim$65 ppm.
%All three values are statistically equivalent.
Table \ref{tab:rms} breaks down the residual rms values by orbit number.  The values given for Orbit~1 do not trend in a consistent direction (larger or smaller) compared to the other orbits.  This suggests that random noise dominates the orbit-to-orbit variations.  As confirmation, the standard error on the residuals in each orbit is $\sim$15~ppm.
%Scatter dominates
%From the observed scatter, it is difficult to draw any conclusions about the 

\begin{table}[t]
    \makebox[\textwidth][c]{
    \begin{tabular}{ccccc|c}
    \hline
    Label       & Orbit 1   & Orbit 2   & Orbit 3   & Orbit 4   & Total  \\
                & (ppm)     & (ppm)     & (ppm)     & (ppm)     & (ppm)  \\
    \hline
    Pre-Flashed & 73        & 62        & 56        & N/A       & 64     \\
    Standard    & 61        & 73        & N/A       & 78        & 72     \\
    %Pre-Flashed & 73{\pm}16 & 62{\pm}16 & 56{\pm}16 & N/A       & 64{\pm}9    \\
    %Standard    & 61{\pm}15 & 73{\pm}14 & N/A       & 78{\pm}14 & 72{\pm}8    \\
    \hline
    \end{tabular}
    }
    \caption{\label{tab:rms}
    \textsl{Measured rms per {\hst} orbit and combined over all orbits.}}
\end{table}

We note that the decrease in rms for the pre-flashed data cannot be attributed to improved centroid stability.  Figure \ref{fig:centroids} demonstrates the increased scatter experienced by the pre-flashed data (relative to the standard data).  Observations with 1$\sigma$ scatter of $\gtrsim$0.1 pixels along the dispersion direction ($x$ axis) experience residual noise due to imprecise flat fielding \citep{ISR-2019-SSMonitor}.  The pre-flashed data exhibit an rms of 0.09 pixels, whereas the standard data exhibit an rms of only 0.04 pixels.  \citet{ISR-2019-SSMonitor} describes how the centroid is computed.

%fig%%%%-------------------------------------------------------%%%%%%
\begin{figure}[tp]
\begin{center}
\mbox{\includegraphics[width=0.8\linewidth]{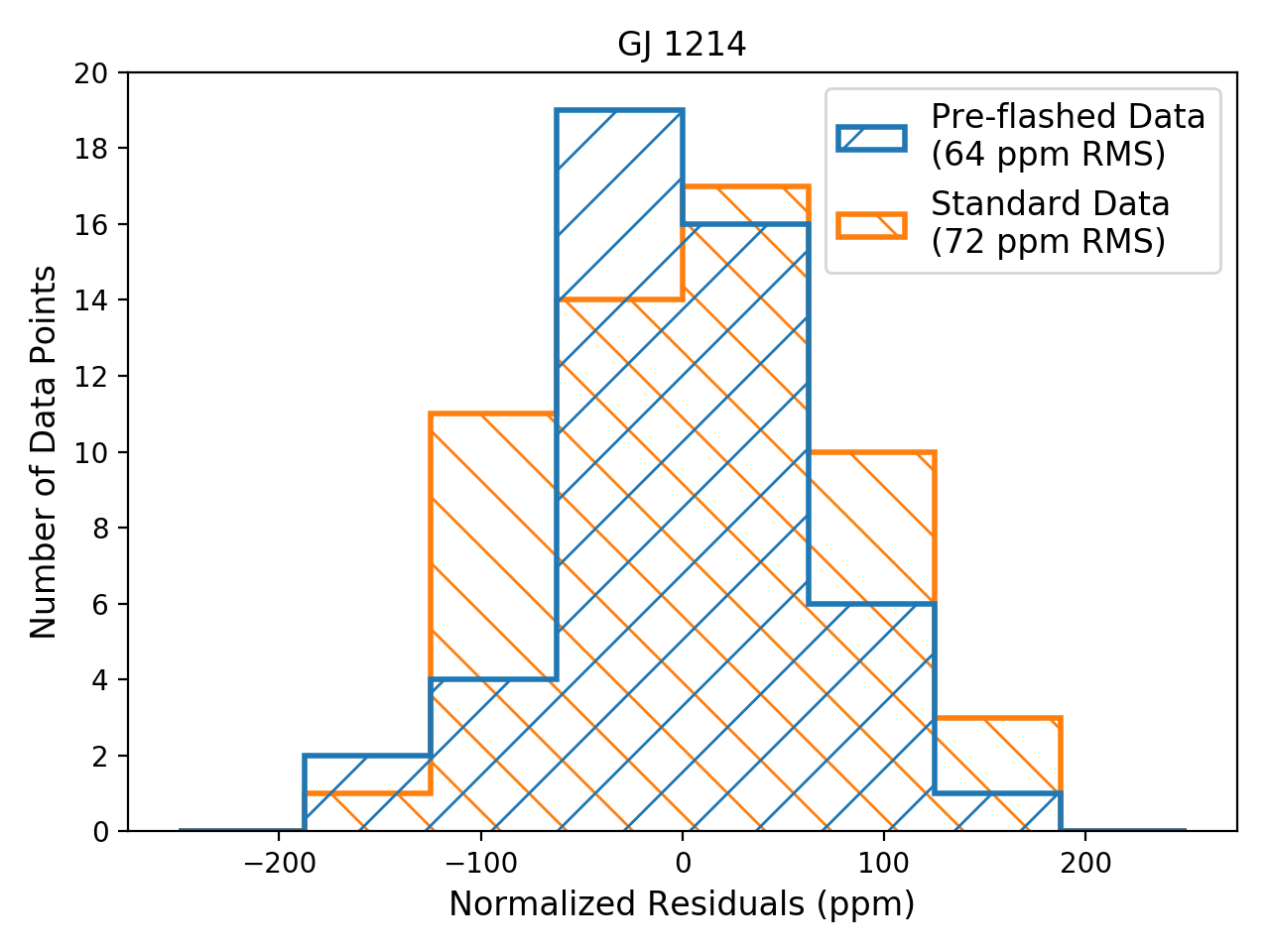}}\vspace{-1.2em}
\caption{\label{fig:res}
\textsl{Histogram of normalized residuals.  The pre-flashed data achieve slightly smaller residual scatter (i.e., narrower histogram) relative to the standard data (64 ppm vs.~72 ppm).  
%Despite having a larger centroid scatter (see Figure \ref{fig:centroids}), .  
The expected photon-limited rms is 65~ppm.}}
\mbox{\includegraphics[width=0.8\linewidth]{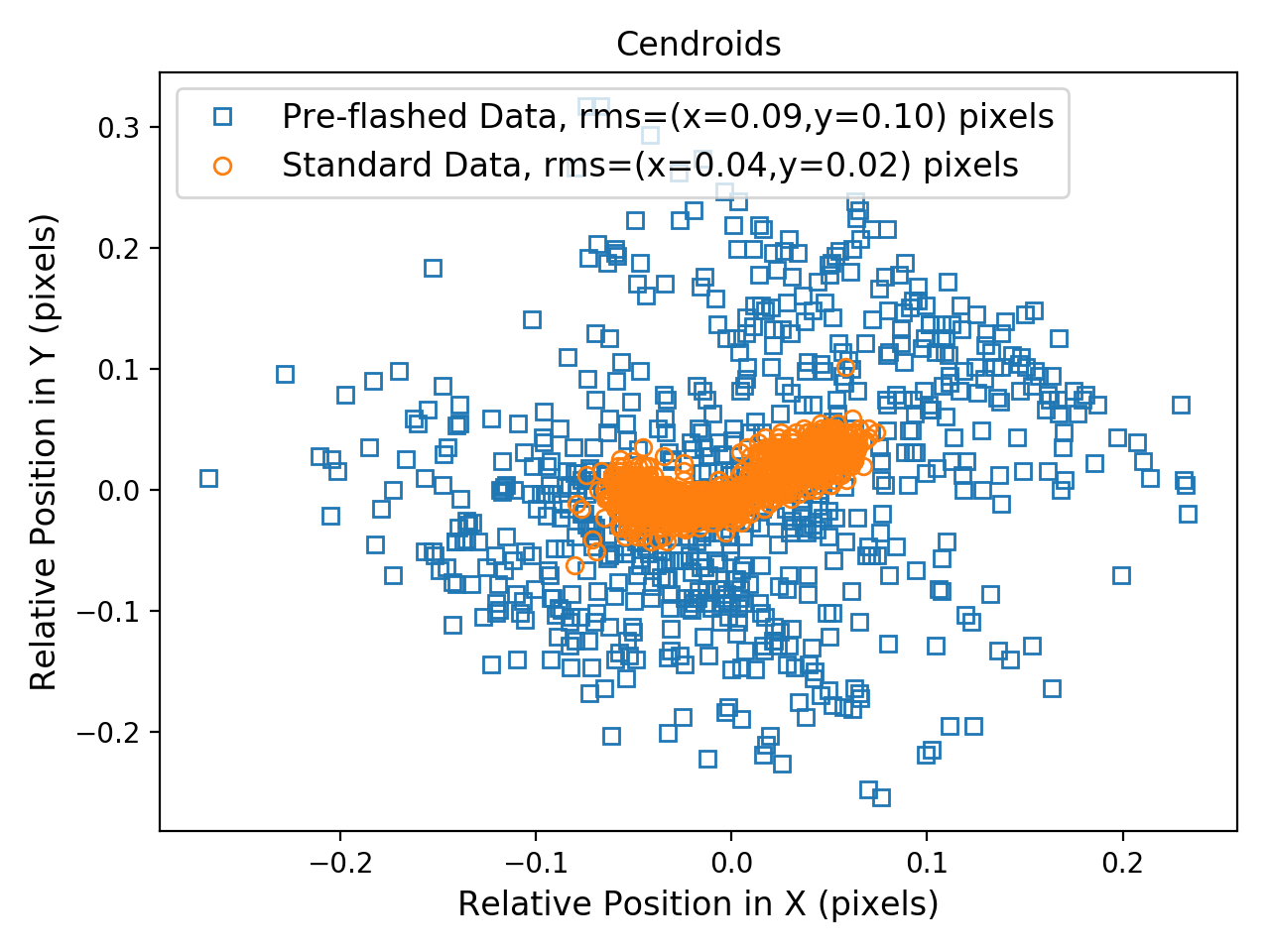}}\vspace{-1.2em}
\caption{\label{fig:centroids}
\textsl{Measured centroid positions of the pre-flashed and standard data.  
%The pre-flashed data's smaller residuals shown in Figure \ref{fig:res} cannot
The increased scatter in the pre-flashed data confirms that the smaller residual scatter shown in Figure \ref{fig:res} is not due to improved centroid stability.
%may have contributed additional noise to the white light curve and corresponding residuals shown in Figure \ref{fig:res}.
}}
\end{center}
\end{figure}
%fig%%%%-------------------------------------------------------%%%%%%

}

%%%%%%%%%%%%%%%%%%%%%%%%%%%%%%%%%%%%%%%%%%%%%%%%%%%%%%%%%%%%%%%%%%%%%%%%%%%%%%%%%%%%%%%%%%%
%                                   Future Implementation
%%%%%%%%%%%%%%%%%%%%%%%%%%%%%%%%%%%%%%%%%%%%%%%%%%%%%%%%%%%%%%%%%%%%%%%%%%%%%%%%%%%%%%%%%%%

%\ssectionstar{Future Implementation}
\normalsize{

When considering whether or not to implement pre-flashing, an important factor is the additional burden on WFC3's channel select mechanism (CSM).  When fully inserted, the CSM directs the light beam into the IR channel.  When the CSM is completely removed, light continues to the UVIS channel.  An intermediate position is necessary to insert the diffuser that directs light from the tungsten lamp into the IR channel.  Thus, the current strategy would typically require two CSM moves per {\hst} orbit, which translates into an additional $\sim$220 moves per year.  This non-negligible increase in wear on the CSM, combined with the effectiveness of current modeling techniques, does not warrant the implementation of pre-flashing on a regular basis.

}

%%%%%%%%%%%%%%%%%%%%%%%%%%%%%%%%%%%%%%%%%%%%%%%%%%%%%%%%%%%%%%%%%%%%%%%%%%%%%%%%%%%%%%%%%%%
%                                   Summary
%%%%%%%%%%%%%%%%%%%%%%%%%%%%%%%%%%%%%%%%%%%%%%%%%%%%%%%%%%%%%%%%%%%%%%%%%%%%%%%%%%%%%%%%%%%

\ssectionstar{Summary}
\normalsize{

We have demonstrated that by pre-flashing the detector during Earth occultation, the magnitude of the ramp is reduced by a factor of $\sim$7 (from 1.30\% to -0.19\%) and, after correcting for instrument systematics using the ramp model described by \citet{Zhou2017}, the improvement in precision is not statistically significant ($\Delta$rms = -8{\pm}9~ppm).  Our strategy of continuously illuminating and reading out the detector (as opposed to holding the charge) validates our hypothesis that the detector more quickly reaches an equilibrium state when subjected to higher levels of flux.  The increase in wear on the CSM and lack of improved precision confirms that the use of pre-flashing in unjustified with WFC3.

%implementing pre-flashing for future time-series observations.

%Since routine implementation of our exact strategy may pose a risk to the health and safety of the WFC3 instrument, we explored alternative strategies that may yield comparable results with less risk.  Future calibration work would be needed to test these strategies and develop a viable path forward.
}

%%%%%%%%%%%%%%%%%%%%%%%%%%%%%%%%%%%%%%%%%%%%%%%%%%%%%%%%%%%%%%%%%%%%%%%%%%%%%%%%%%%%%%%%%%%
%                                  Appendix
%%%%%%%%%%%%%%%%%%%%%%%%%%%%%%%%%%%%%%%%%%%%%%%%%%%%%%%%%%%%%%%%%%%%%%%%%%%%%%%%%%%%%%%%%%%
\ssectionstar{Appendix A}

Successfully implementing this program required several iterations with the scheduling team that we record here for posterity.  When initially submitted, the program contained a number of timing and orient constraints that were incompatible with the scheduling software. Specifically, Visit 01 (the first pre-flash orbit) had separate orient ranges, despite being internal, as well as a SAME ORIENT restriction linking it to Visits 02 -- 04. It also had both GROUP WITHIN and PHASE requirements. The combination of all these restrictions crashes the scheduling software.  After testing with the Spike team (Spike is a general framework for planning and scheduling {\hst} observations), it became clear that visits that do not have an external target cannot have Orient constraints placed upon them.  In order to fix this, all Orient ranges need to be placed on one of the ``pointed'' visits (Visit 02). The unpointed visit (Visit 01) can then use the SAME ORIENT AS Visit 02 restriction.  For the same reason, the phase constraint must also be placed on the pointed visit.

It was a requirement of this program that the first lamp exposure take place during occultation immediately preceding the first pointed visit. To implement this requirement, we added a timing link to Visit 02 to be After Visit 01 by 30-45 minutes. Since Visit 02 has to execute at the start of its orbit, this requirement forces Visit 01 to take place half an hour beforehand, which is during Earth occultation.

We briefly examined if the pre-flashing strategy could be implemented as a calibration program that runs parallel to time-series, spatial scan observations and determined that it would likely be feasible.  Having the pre-flashing program run in parallel would minimize any additional work performed by observers when planning their Phase II programs.  A similar strategy was used to schedule darks after South Atlantic Anomaly crossings with NICMOS \citep{ISR-NICMOS-2003-10}.

%%%%%%%%%%%%%%%%%%%%%%%%%%%%%%%%%%%%%%%%%%%%%%%%%%%%%%%%%%%%%%%%%%%%%%%%%%%%%%%%%%%%%%%%%%%
%                                  Acknowledgements
%%%%%%%%%%%%%%%%%%%%%%%%%%%%%%%%%%%%%%%%%%%%%%%%%%%%%%%%%%%%%%%%%%%%%%%%%%%%%%%%%%%%%%%%%%%
\ssectionstar{Acknowledgements}

We would like to thank Jules Fowler, Peter McCullough, Sylvia Baggett, and John Mackenty for their thorough reviews of this ISR.  We would also like to thank Mario Gennaro and Knox Long for their useful insight and discussions on this program.

%%%%%%%%%%%%%%%%%%%%%%%%%%%%%%%%%%%%%%%%%%%%%%%%%%%%%%%%%%%%%%%%%%%%%%%%%%%%%%%%%%%%%%%%%%%
%                                            Reference List
%%%%%%%%%%%%%%%%%%%%%%%%%%%%%%%%%%%%%%%%%%%%%%%%%%%%%%%%%%%%%%%%%%%%%%%%%%%%%%%%%%%%%%%%%%%
%\newpage

%\begin{multicols}{2}
{%\small \footnotesize
\bibliography{ms}
}
%\end{multicols}

%%%%%%%%%%%%%%%%%%%%%%%%%%%%%%%%%%%%%%%%%%%%%%%%%%%%%%%%%%%%%%%%%%%%%%%%%%%%%%%%%%%%%%%%%%%
%                                        End of Document
%%%%%%%%%%%%%%%%%%%%%%%%%%%%%%%%%%%%%%%%%%%%%%%%%%%%%%%%%%%%%%%%%%%%%%%%%%%%%%%%%%%%%%%%%%%
\clearpage
\end{document}

%% file: kbsmacs.tex
\newcommand\degree{\degr}
\newcommand\degrees\degree

\newcommand\hst{\em HST}

\DeclareSymbolFont{UPM}{U}{eur}{m}{n}
\DeclareMathSymbol{\umu}{0}{UPM}{"16}
\let\oldumu=\umu
\renewcommand\umu{\ifmmode\oldumu\else\math{\oldumu}\fi}
\newcommand\micro{\umu}
\renewcommand\micron{\micro m}
\newcommand\microns \micron

\renewcommand\arcsec[0]{$^{\prime\prime}$}

\let\oldsim=\sim
\renewcommand\sim{\ifmmode\oldsim\else\math{\oldsim}\fi}
\let\oldpm=\pm
\renewcommand\pm{\ifmmode\oldpm\else\math{\oldpm}\fi}
\newcommand\by{\ifmmode\times\else\math{\times}\fi}

\newbox{\wdbox}
\renewcommand\c{\setbox\wdbox=\hbox{,}\hspace{\wd\wdbox}}
\renewcommand\i{\setbox\wdbox=\hbox{i}\hspace{\wd\wdbox}}

\newcount\timect
\newcount\hourct
\newcount\minct
\newcommand\now{\timect=\time \divide\timect by 60
         \hourct=\timect \multiply\hourct by 60
         \minct=\time \advance\minct by -\hourct
         \number\timect:\ifnum \minct < 10 0\fi\number\minct}

\catcode`@=11

\newcommand\comment[1]{}
\newcommand\commenton{\catcode`\%=14}
\newcommand\commentoff{\catcode`\%=12}

\renewcommand\math[1]{$#1$}
\newcommand\mathshifton{\catcode`\$=3}
\newcommand\mathshiftoff{\catcode`\$=12}

\comment{the backslash is necessary}

\comment{alignment tab}

\let\atab=&
\newcommand\atabon{\catcode`\&=4}
\newcommand\ataboff{\catcode`\&=12}

\let\oldmsp=\sp
\let\oldmsb=\sb
\def\sp#1{\ifmmode
           \oldmsp{#1}%
         \else\strut\raise.85ex\hbox{\scriptsize #1}\fi}
\def\sb#1{\ifmmode
           \oldmsb{#1}%
         \else\strut\raise-.54ex\hbox{\scriptsize #1}\fi}
\newbox\@sp
\newbox\@sb
\def\sbp#1#2{\ifmmode%
           \oldmsb{#1}\oldmsp{#2}%
         \else
           \setbox\@sb=\hbox{\sb{#1}}%
           \setbox\@sp=\hbox{\sp{#2}}%
           \rlap{\copy\@sb}\copy\@sp
           \ifdim \wd\@sb >\wd\@sp
             \hskip -\wd\@sp \hskip \wd\@sb
           \fi
        \fi}
\def\msp#1{\ifmmode
           \oldmsp{#1}
         \else \math{\oldmsp{#1}}\fi}
\def\msb#1{\ifmmode
           \oldmsb{#1}
         \else \math{\oldmsb{#1}}\fi}
\def\supon{\catcode`\^=7}
\def\supoff{\catcode`\^=12}
\def\subon{\catcode`\_=8}
\def\suboff{\catcode`\_=12}
\def\supsubon{\supon \subon}
\def\supsuboff{\supoff \suboff}

\newcommand\actcharon{\catcode`\~=13}
\newcommand\actcharoff{\catcode`\~=12}

\newcommand\paramon{\catcode`\#=6}
\newcommand\paramoff{\catcode`\#=12}

\comment{And now to turn us totally on and off...}

\newcommand\reservedcharson{\commenton \mathshifton \atabon \supsubon \actcharon
	\paramon}

\newcommand\reservedcharsoff{\commentoff \mathshiftoff \ataboff
	\supsuboff \actcharoff \paramoff}

\catcode`@=12
\reservedcharsoff

\reservedcharson

\comment{ Must have ONLY ONE of these... trust these macros, they work

}

\newcommand{\squishlist}{
 \begin{list}{$\bullet$}
  { \setlength{\itemsep}{1pt}
     \setlength{\parsep}{0pt}
     \setlength{\topsep}{3pt}
     \setlength{\partopsep}{0pt}
     \setlength{\leftmargin}{2.0em}
     \setlength{\labelwidth}{1.5em}
     \setlength{\labelsep}{0.5em} } }

\newcommand{\squishend}{
  \end{list}  }

\reservedcharson
\actcharon